\documentclass[useAMS,usenatbib]{mn2e} 
\usepackage{aas_macros}
\usepackage{graphics}
\usepackage{epsfig}  
\usepackage{natbib} 
\usepackage{float}
\newcommand{\amiga}{\texttt{AMIGA}}
\newcommand{\ahf}{\texttt{AHF}}

\newcommand{\mlapm}{\texttt{MLAPM}}
\newcommand{\mhf}{\texttt{MHF}}

\newcommand{\Vec}[1]{{\bf #1}}
\newcommand{\Tab}[1]{Table~\ref{#1}}
\newcommand{\Sec}[1]{Section~\ref{#1}}
\newcommand{\Eq}[1]{Eq.~(\ref{#1})}
\newcommand{\Fig}[1]{Figure~\ref{#1}}
\newcommand{\hMpc}{{\ifmmode{h^{-1}{\rm Mpc}}\else{$h^{-1}$Mpc}\fi}}
\newcommand{\hkpc}{{\ifmmode{h^{-1}{\rm kpc}}\else{$h^{-1}$kpc}\fi}}
\newcommand{\hMsun}{{\ifmmode{h^{-1}{\rm {M_{\odot}}}}\else{$h^{-1}{\rm{M_{\odot}}}$}\fi}}

\def\lesssim{\mathrel{\hbox{\rlap{\hbox{\lower4pt\hbox{$\sim$}}}\hbox{$<$}}}}
\def\gtrsim{\mathrel{\hbox{\rlap{\hbox{\lower4pt\hbox{$\sim$}}}\hbox{$>$}}}}
\def\LCDM{$\Lambda$CDM}

\title[Radial alignment and host mass]{On the relation between radial
  alignment of dark matter subhalos and host mass in cosmological
  simulations} \author[Knebe et al.] {Alexander Knebe$^1$, Nadya
  Draganova$^1$, Chris Power$^2$, Gustavo Yepes$^3$, 
  \newauthor Yehuda Hoffman$^4$, Stefan Gottl\"ober$^1$, and Brad K. Gibson$^5$
  \\
  $^1$Astrophysical Institute Potsdam, An der Sternwarte 16, Germany
  \\
  $^2$Department of Physics \& Astronomy, University of Leicester,
  University Road, Leicester LE1 7RH, UK
  \\
  $^3$Grupo de Astrof\'{\i}sica, Universidad Aut\'onoma de Madrid,
  Madrid E-28049, Spain
  \\
  $^4$Racah Institute of Physics, Hebrew University, Jerusalem 91904,
  Israel
  \\
  $^5$Centre for Astrophysics, University of Central Lancashire, Preston PR1 2HE, UK
  }

\begin{document}

\date{Submitted Version ...}

\pagerange{\pageref{firstpage}--\pageref{lastpage}} \pubyear{2008}

\maketitle

\label{firstpage}

\begin{abstract}
  We explore the dependence of the radial alignment of subhalos on the
  mass of the host halo they orbit in. As the effect is seen on a
  broad range of scales including massive clusters as well as galactic
  systems it only appears natural to explore this phenomenon by means
  of cosmological simulations covering the same range in masses. We
  have 25 well resolved host dark matter halos at our disposal ranging
  from $10^{15}$\hMsun\ down to $10^{12}$\hMsun\ each consisting of
  order of a couple of million particles within the virial radius. We
  observe that subhalos tend to be more spherical than isolated
  objects. Both the distributions of sphericity and triaxiality of
  subhalos are Gaussian distributed with peak values of $\langle s
  \rangle \approx 0.80$ and $\langle T \rangle \approx 0.56$,
  irrespective of host mass. Interestingly we note that the radial
  alignment is independent of host halo mass and the distribution of
  $\cos\theta$ (i.e. the angle between the major axis $E_a$ of each
  subhalo and the radius vector of the subhalo in the reference frame
  of the host) is well fitted by a simple power law $P(\cos\theta)
  \propto \cos^4\theta$ with the same fitting parameters for all host
  halos.
\end{abstract}

\begin{keywords}
  galaxies: evolution -- galaxies: haloes -- cosmology: theory --
  cosmology: dark matter -- methods: $N$-body simulations
\end{keywords}

\section{Introduction}
\label{sec:introduction}

The concordance of multitude of recent cosmological studies has
demonstrated that we appear to live in a spatially flat,
$\Lambda$-dominated cold dark matter ($\Lambda$CDM) universe
\cite[cf. ][]{Spergel07}. During the past decade simulation codes and
computer hardware have advanced to such a stage where it has been
possible to resolve in detail the formation of dark matter halos and
their subhalo populations in a cosmological context
\citep[e.g. ][]{Klypin99}. These results, coupled with the
simultaneous increase in observational data (e.g. 2 degree Field
galaxy redshift survey (2dFGRS), \citet{2dF-DR}; Sloan Digital Sky
Survey (SDSS), \citet{SDSS-DR5}), has opened up a whole new window on
the concordance cosmogony in the field that has become known as
``near-field cosmology'' \citep{Freeman02}.

One property of the galaxy population that has been measured is the
spatial distribution of satellites about their primaries. Some
evidence suggests that this is anisotropic, with satellites clustering
about the major axis of the primary. An anisotropic spatial
distribution has been measured in various observational data sets
\citep[e.g. ][please note that its interpretation continues to be a
matter of debate though]{Zaritsky97, Sales04, Brainerd05, Yang06,
  Bailin07}; a similar result is obtained for the distribution of
subhalos in simulations of cosmological dark matter halos
\citep[e.g. ][]{Knebe04, Zentner05, Libeskind05}, as well as following
from theoretical modelling \citep{Lee05}.

Another property of the satellite population that has been measured is
the radial alignment of their primary axes with respect to the
host. The first evidence for this effect was reported for the Coma
cluster, where it was observed that the projected major axes of
cluster members preferentially align with the direction to the cluster
centre \citep{Hawley75, Thompson76}.  Such a correlation between
satellite elongation and radius vector has further been confirmed by
statistical analysis of the SDSS data \citep{Pereira05, Agustsson06,
  Faltenbacher07a, Wang07}. However, we also acknowledge that
\citet{Bernstein02} did not find such a signal in the 2dFGRS.  The
radial alignment of subhalo shapes towards the centre of their host
has also been measured for the subhalo population in cosmological
simulations \citep{Kuhlen07, Faltenbacher07b, Pereira07}. We note that
this effect was predicted by \citet{Ciotti94}, who used simulations to
argue that cluster galaxies were influenced by the tidal field of the
host cluster tidal field.

In this \textit{Letter} we provide evidence that the radial alignment of
subhalos in cosmological simulations does not depend on the mass of their
host dark matter halo. In addition, we show that subhalo triaxialities and
shapes follow a Gaussian distribution, which again does not depend on the mass
of the host.

\section{The Data}
\label{sec:hosts}

\begin{table}
\begin{center}
  \caption{Details about the hosts and their subhalo populations.
    $N_{p, \rm host}$ gives the number of particles in the host while
    $M_{\rm vir}$ measures its mass (in $10^{14}$\hMsun) and the mass
    $R_{\rm host}$ its virial radii (in \hkpc).  The last two columns
    give the number of satellites in excess of $N_{p}>200$ and
    $N_{p}>200, b/a<0.9$, respectively. These (sub-)samples of all
    identified subhalos comply with the criteria to reliably measure
    a) shape (S) and b) radial alignment (RA). Please refer to the
    text for further details.}
\begin{tabular}{cccccc}
\hline
host  & $N_{p, \rm host}$      &  $M_{\rm host}$      & $R_{\rm host}$ & $N_{\rm sub}^{\rm S}$ & $N_{\rm sub}^{\rm RA}$ \\
\hline
MC1  &  2608898 &  13.05 &    2288 &    217 &    115  \\
MC2  &  2531440 &  10.90 &    2155 &    165 &     79  \\
MC3  &  2275913 &  15.37 &    2417 &    137 &     77  \\
C1  &  1764131  &   2.87 &    1355 &    85 &     33  \\
C2  &  864068  &   1.41 &    1067 &    61 &     33  \\
C3  &  654169  &   1.06 &     973 &    27 &     11  \\
C4  &  859385  &   1.40 &    1061 &    38 &     12  \\
C5  &  725694  &   1.18 &    1008 &    34 &     10  \\
C6  &  869614  &   1.41 &    1065 &    46 &     17  \\
C7  &  1752783  &   2.85 &    1347 &    90 &     30  \\
C8  &  1868533  &   3.05 &    1379 &    127 &     56  \\

C9   &  1755223 &   1.25 &    1047 &    112 &     53  \\
C10  &  1918720 &   1.01 &     976 &    91 &     54  \\
C11  &  1918359 &   1.59 &    1133 &    117 &     68  \\
C12  &  894831 &   1.82 &    1187 &    80 &     34  \\
G1   &  1094732 &    0.178 &     547 &    59 &     46  \\
G2   &  1093805 &    0.178 &     547 &    90 &     46  \\
G3   &  1083392 &    0.176 &     545 &    97 &     48  \\
G4   &  1010659 &    0.164 &     532 &    48 &     32  \\
G5   &  971312 &    0.158 &     525 &    42 &     34  \\
MW1 &   2226368  & 0.014 &     227 &    100 &     28  \\
MW2 &   1734197  & 0.019 &     250 &    81 &     66  \\
MW3 &  1535393  & 0.017 &     242 &    86 &     49  \\
MW4 &   1384565  & 0.014 &     228 &    73 &     40  \\
MW5 &   1155977  & 0.022 &     265 &    71 &     25 
\end{tabular}
\label{tab:hosts}
\end{center}
\end{table}

\subsection{The Host Halos}
We use a set of 25 high-resolution cosmological (zoom) simulations of
individual dark matter host halos. Three of these runs (MC1-3)
represent massive galaxy cluster of $M\approx 10^{15}$\hMsun. We
have twelve cluster-sized objects (C1-12) with $M\approx
10^{14}$\hMsun, there are five group-sized systems (G1-5) with
$M\approx 10^{13}$\hMsun, and another five Milky Way-type halos
(MW1-5) with $M\approx 10^{12}$\hMsun. The particulars of these hosts
are summarised in \Tab{tab:hosts}. 

For the interested reader, C1-8 have been generated with the adaptive
mesh refinement code \texttt{MLAPM} \citep{Knebe01} and their
properties have been discussed in detail in \citet{Warnick06} and
\citet{Warnick08}. MW1 corresponds to model ``Box20'' of
\citet{Prada06} that has been simulated with the \texttt{ART} code
\citep{Kravtsov97}. All other halos (i.e. MC1-3, C9-12, G1-5, MW2-5)
have been simulated with the Tree-PM code \texttt{GADGET2}
\citep{Springel05} and their detailed properties will be presented in
a forthcoming paper (Knollmann et al., in preparation). However, we
wish to highlight that G1-G5 are drawn from a constrained realisation
of the Local Universe ($1024^3$ particles in a box of side 64\hMpc;
Yepes et al., in preparation).\footnote{See \citet{Klypin03} for a
  presentation of the method applied to run constrained simulations.}
MC1-3 are resimulations of individual halos embedded in a box of side
512\hMpc; MW2-4 are resimulations of individual halos embedded in a
box of side 50\hMpc, while MW5 is a resimulation embedded in a box of
side 150\hMpc.

Note that MC1-3, C9-12 and G1-5 have been simulated assuming a WMAP3 cosmology
\citep{Spergel07}, while all other systems were simulated assuming a WMAP1
cosmology \citep{Spergel03}.

\subsection{The Subhalos}
Both host halos and subhalos are identified using
\ahf\footnote{{\small \textbf{A}MIGA}'s-{\small
    \textbf{H}}alo-{\small\textbf{F}}inder; \ahf\ can be downloaded
  from \texttt{http://www.aip.de/People/aknebe/AMIGA}. \amiga\ is the
  successor to \mlapm.} (Knollmann et al., in preparation), an MPI parallelised
modification of the
\mhf\footnote{\mlapm's-\texttt{H}alo-\texttt{F}inder} algorithm
presented in \cite{Gill04a}. \ahf~utilises an adaptive grid hierarchy
to locate halos (subhalos) as peaks in an adaptively smoothed density
field. Local potential minima are computed for each peak and the set
of particles that are gravitationally bound to the peak are
returned. If the peak contains in excess of 20 particles, then it is
considered a halo (subhalo) and retained for further analysis.

For each halo (subhalo) we calculate a suite of canonical properties
from particles within the virial (truncation) radius. We define the
virial radius $R_{\rm vir}$ as the point at which the density profile
(measured in terms of the cosmological background density $\rho_b$)
drops below the virial overdensity $\Delta_{\rm vir}$, i.e. $M(<R_{\rm
  vir})/(4\pi R_{\rm vir}^3/3) = \Delta_{\rm vir}
\rho_b$.\footnote{For a distinct (i.e. host) halo in a \LCDM\
  cosmology with the cosmological parameters that we have adopted,
  $\Delta_{\rm vir}=340$ at $z=0$.} This prescription is not
appropriate for subhalos in the dense environs of their host halo,
where the local density exceeds $\Delta_{\rm vir} \rho_b$, and so the
density profile will show a characteristic upturn at a radius $R
\lesssim R_{\rm vir}$. In this case we use the radius at which the
density profile shows this upturn to define the truncation radius for
the subhalo. Further details of this approach can be found in
\cite{Gill04a}.

\section{Subhalo Shapes}
\label{sec:shape}
A generic prediction of the CDM model is that dark matter halos are
triaxial systems, that can be reasonably approximated as ellipsoids
\citep[e.g.][]{Frenk88, Warren92, Kasun05, Bailin05, Allgood06,
  Maccio07, Bett07, Gottloeber07}.  Following others
\citep[e.g.][]{Gerhard83, Bailin05, Allgood06, Pereira07, Kuhlen07,
  Faltenbacher07b}, we measure the shape of the halos by the weighted
moment of inertia:

\begin{equation}
I_{ij} = \sum_{k}m_{k}\frac{r_{ki}r_{kj}}{r_{k}^{2}}.
\end{equation}

\noindent
The axis ratios $b/a$ and $c/a$ are the square roots of the
eigenvalues ($a>b>c$) and the corresponding eigenvectors $\Vec{E}_a,
\Vec{E}_b, \Vec{E}_c$ give the directions of the principal axes. 


We wish to measure the direction of a subhalo's principal axes with
respect to the centre of the host, and so it is important to measure
reliably shape and orientation. Therefore we follow \citet{Pereira07}
and require subhalos to contain at least $N_{p, \rm min}=200$ for
reliable shape estimation in this Section, and to have an axis ratio
$b/a<0.9$ when investigating the radial alignments in
\Sec{sec:alignments}.

The numbers of subhalos in each host halo compliant with either of
these two criteria are given in \Tab{tab:hosts}.  One observes that
there exists a prominent population of subhalos with $b/a >
0.9$. Careful inspection reveals that the distribution of $P(b/a)$
(not shown here) is Gaussian with the peak at about $0.89$. This means
that there is a substantial number of objects orbiting within the
virial radius of the host which do not enter into our radial alignment
analysis below. However, as also noted by \citet{Pereira07}, we
confirm that the majority of these systems is spherical and hence a
major axes cannot be robustly defined.\footnote{If we nevertheless
  include objects with $b/a \geq 0.9$ in the radial alignment analysis
  the signal is practically unaffected.}

Because we are interested in the dependence of both the shape and
the radial alignment on the mass of the host, we stack data accordingly. 
All subhalos for MC, C, G, and MW hosts are combined into one single plot 
and hence there appear four panels in subsequent plots, one for each host 
class.

\subsection{Shape Measurements}

\paragraph*{Triaxiality} 
Using the eigenvalues $a > b > c$ of the moment of inertia tensor we
calculate the triaxiality parameter \citep[e.g. ][]{Franx91}

\begin{equation}
T = \frac{a^{2}-b^{2}}{a^{2}-c^{2}} \ .
\end{equation}

\begin{figure}
\centering{\epsfig{file=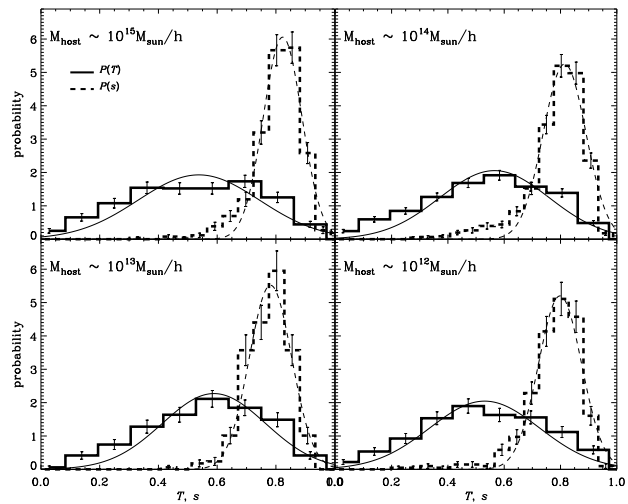,width=85mm}}
\caption{Distribution of subhalo triaxialities and subhalo shapes. Only
         subalos with more than $N_{p}>200$ are considered.
         \label{fig:Pshape}}
\end{figure}

\noindent
In \Fig{fig:Pshape} we show the triaxiality probability distribution
(solid histograms showing the fraction of subhalos in the respective
bin and normalized to unity) for subhalos containing more than 200
particles; error bars assume Poisson errors. We further fitted a
Gaussian to $P(T)$ and list the two fitting parameters (width $\sigma$
and peak $T_{\rm peak}$) in \Tab{tab:shapes}.

\paragraph*{Sphericity} 
Although triaxiality is useful for distinguishing prolate from oblate
halos, we seek a more reliable measure for deviation from
sphericity. We use the axis ratio

\begin{equation}
 s = \frac{c}{a}
\end{equation}

\noindent
for this purpose; a spherical halo has $c/a \rightarrow 1$ while a
highly oblate or prolate halo has $c/a \rightarrow 0$. The resulting
distributions for subhalos in excess of 200 particles can be viewed in
\Fig{fig:Pshape} (dashed histograms), too. The corresponding best-fit
parameters of a Gaussian are again listed in \Tab{tab:shapes}.

\begin{table}
\begin{center}
\caption{Gaussian's parameters for the probability distribution of triaxiality and sphericity.} 
\begin{tabular}{ccccccc}
\hline
host & & \multicolumn{2}{c}{$P(T)$} & &  \multicolumn{2}{c}{$P(s)$}\\
     & &     $T_{\rm peak}$ & $\sigma$     & &        $s_{\rm peak}$ &  $\sigma$  \\ \hline
 MC  & &       0.54  &   0.21       & &         0.82  &     0.07   \\ 
 C   & &       0.57  &   0.19       & &         0.81  &     0.08   \\
 G   & &       0.59  &   0.18       & &         0.78  &     0.07   \\
 MW  & &       0.53  &   0.20       & &         0.80  &     0.08   
\end{tabular}
\label{tab:shapes}
\end{center}
\end{table}

\subsection{Discussion}
There is an extensive literature on the shape of isolated/field dark
matter halos \citep[e.g. ][]{Frenk88, Warren92, Kasun05, Bailin05,
  Allgood06, Maccio07, Bett07}. These studies indicate that the
average triaxiality and sphericity of isolated halos is $\langle
T\rangle \approx 0.75$ and $\langle s\rangle\approx 0.66$,
respectively.

The first extension of such studies to subhalo populations was
presented by \citet{Kuhlen07} for the ``Via Lactea'' simulation
\citep{Diemand07}, an ultra-high resolution simulation of a Milky
Way-type dark matter halo. We do not have the resolution of the Via
Lactea simulation but we do have a statistical sample of sufficiently
well resolved systems to explore the shapes of subhalos.

In agreement with \citet{Kuhlen07}, we observe that subhalos are
triaxial with $\langle T \rangle \approx 0.55$ (cf. \Tab{tab:shapes}),
irrespective of the host mass. We note also that the distribution of
$T$ as presented in \Fig{fig:Pshape} (solid histograms) is well fitted
by a Gaussian. The same holds for the distribution of sphericities
$P(s)$ presented in \Fig{fig:Pshape} (dashed histograms), too
(cf. also Fig.3 in \citet{Faltenbacher07b}). However, our sphericities
are typically $\langle s \rangle \approx 0.80$ and are marginally
larger than the ones reported by \citet{Kuhlen07} and
\citet{Faltenbacher07b} ($\langle s \rangle \approx 0.74$). This
likely reflects the different halo finding algorithms; both
\citet{Kuhlen07} and \citet{Faltenbacher07b} use a method derived from
the friends-of-friends algorithm, whereas we use a spherical halo
finder \texttt{AHF} (cf. \citet{Gill04a} who performed an in-depth
comparison of \texttt{AHF}, FOF and SKID.)

\section{Subhalo Alignments}
\label{sec:alignments}
The principal aim of this study is to investigate whether or not there
is a dependence of the radial alignment of subhalos (i.e. the
alignment of their major axis with respect to the centre of the host)
on the mass of their host halo. Recent observational evidence suggests
that the major axis (in projection) of satellite galaxies tend to
``point towards the centre of their host'' \citep[e.g. ][]{Pereira05,
  Agustsson06, Yang06, Faltenbacher07a, Wang07}. It is therefore
natural to ask whether or not subhalos in cosmological simulations
display a similar trend. To date a few studies have investigated this
subject \citep{Kuhlen07, Faltenbacher07b, Pereira07}. We extend this
work by examining the mass dependence of the radial alignment.

\subsection{Measurement of Radial Alignment}
To measure the radial alignment of subhalos, we use the
eigenvector $\Vec{E}_a$ which corresponds to the direction of the
major axis $a$ of the subhalo. We quantify the radial alignment of subhalos 
as the angle between the major axis $E_a$ of each subhalo and the radius
vector of the subhalo in the reference frame of the host:

\begin{equation}
\cos \theta = \frac{\Vec{R}_{\rm sub}*\Vec{E}_{a, \rm sub}}{|\Vec{R}_{\rm sub}| |\Vec{E}_{a, \rm sub}|}
\end{equation}

\begin{figure}
\centering{\epsfig{file=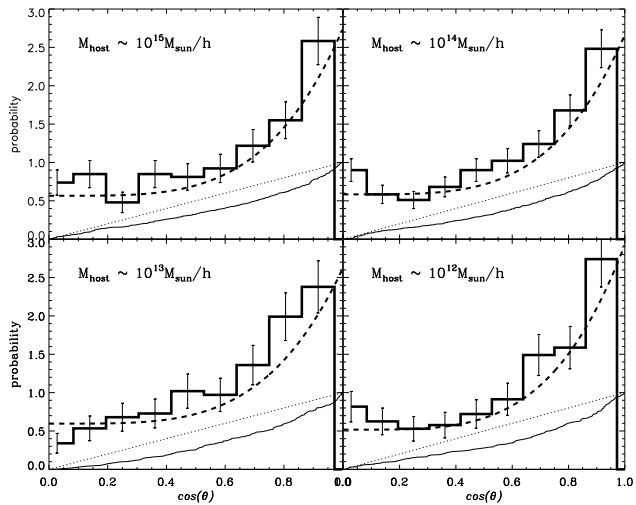,width=85mm}}
\caption{Distribution of subhalo radial alignment. The histograms are
  the differential distribution (with Poissonian error bars) that have
  been fitted by \Eq{eq:PcosREsat} (dashed line). The continuous line
  represents the cumulative probability distribution
  $P(<\hspace*{-0.6mm}\cos\theta)$; the dotted line is the
  (cumulative) isotropic distribution. Only subhalos with $N_{p}>200$ 
  and $b/a<0.9$ are considered.
 \label{PcosREsat}}
\end{figure}

\noindent
The (normalized) distribution $P(\cos\theta)$ of $\cos\theta$
measuring the fraction of subhaloes in the respective bin can be
viewed in \Fig{PcosREsat}.

We find a positive radial alignment signal different from isotropy, in
agreement with \citet{Kuhlen07}, \citet{Faltenbacher07b} and
\citet{Pereira07}. This can be verified in \Tab{tab:PcosREsat} where
we compare the cumulative probability distributions
$P(<\hspace{-0.5mm} \cos\theta)$ (shown as thin solid lines in
\Fig{PcosREsat}, too) with the isotropic distribution by applying a
Kolmogorov-Smirnoff (KS) test. The resulting KS probabilities are
consistent with zero. They are listed in \Tab{tab:PcosREsat} alongside
$D$, the maximum distance of the actual and isotropic distribution.

To better quantify these differences, we fit the following heuristically
determined function to the (differential) probability distributions

\begin{equation} \label{eq:PcosREsat}
 P(x) = \left(\frac{1}{B+A/5}\right) A x^4 + B \ ,
\end{equation}

\noindent
with $x=\cos\theta$. In order to gauge the error introduced by binning
the data we perform the fit to differently binned $P(\cos\theta)$
changing $N_{\rm bins}$ steadily from 5 to 15; $\langle A\rangle$ and
$\langle B\rangle$ given in \Tab{tab:PcosREsat} are the corresponding
means of the best-fit parameters and the errors are the
$1\sigma$-deviation. We note that these values are practically
identical across hosts and do not depend on mass. This suggests that
there is no (hidden or obvious) relation between the radial alignment
of the subhalos and the mass of the host. This implies that any
explanation of this phenomenon has to apply to galactic halos as well
as clusters of mass $10^{15}$\hMsun.

\begin{table}
\begin{center}
  \caption{Kolmogorov-Smirnov test and best fit parameters for radial
    alignment probability distribution. The errors for $A$ and $B$ are
    based upon fitting \Eq{eq:PcosREsat} to $P(\cos\theta)$ using
    different number of bins $N_{\rm bins} \in [5,15]$ and $\langle A
    \rangle$ and $\langle B \rangle$ are the respective mean values.}
\begin{tabular}{ccccc}
\hline
host & $D$ &  KS probability & $\langle A\rangle$ & $\langle B \rangle$ \\ \hline

MC  &  0.266 & $6\times 10^{-09}$ & 2.81 $\pm$ 0.22 & 0.56 $\pm$ 0.05 \\ 
 C  &  0.262 & $6\times 10^{-13}$ & 2.67 $\pm$ 0.23 & 0.60 $\pm$ 0.05 \\
 G  &  0.306 & $5\times 10^{-09}$ & 2.50 $\pm$ 0.25 & 0.60 $\pm$ 0.04 \\
 MW &  0.327 & $3\times 10^{-10}$ & 2.58 $\pm$ 0.26 & 0.58 $\pm$ 0.04
\end{tabular}
\label{tab:PcosREsat}
\end{center}
\end{table}

\subsection{Discussion}

Both \citet{Yang06} and \citet{Wang07} report that the spatial anisotropy of
satellite galaxies depends on host halo mass. In contrast, no such mass
dependence has been reported for the radial alignment of satellite galaxies
with respect to the host. This can be understood if one considers the physical
origin of these effects. The spatial anisotropy is linked to the anisotropic
infall of subhalos (and presumably satellite galaxies) onto their host halo
\citep[e.g. ][]{Knebe04}, and so can be considered an environmental effect. In
contrast, the radial alignment is a dynamical effect \citep{Kuhlen07, 
  Pereira07}, driven by the tidal field of the host halo \citep{Ciotti94}; the
subhalo adjusts its orientation in response to the tidal field on a timescale
that is much shorter than the Hubble time. However, it is difficult to
determine whether or not a subhalo adjusts its orientation by rigid body 
rotation \citep[e.g.][]{Kuhlen07, Pereira07} or by changing its shape; it 
is not straightforward to distinguish between changes in subhalo shape and 
``simple'' figure rotation, especially for subhalos which are suffering mass
loss. While the latter has been confirmed for isolated/host halos 
\citep{Bailin04} it has yet to be verified for subhalos.

\section{Summary and Conclusions}
\label{sec:summary}

We have examined the whether or not the radial alignment of the major
axes of subhalos with respect to the centre of their host dark matter
halo is dependent on host halo mass. Our results draw upon a sample of
25 cosmological high resolution resimulations of halos spanning
galaxy- to cluster-mass scales ($10^{12}$\hMsun to $10^{15}$\hMsun),
the majority of them containing in excess of $10^6$ particles within
the virial radius. Our main results may be summarised as follows:

\begin{itemize}

\item Subhalos tend to be more spherical than isolated halos, with $\langle s
  \rangle \approx 0.80$ .
 
\item Subhalos have triaxialities $\langle T \rangle \approx 0.53$,
  lower than isolated halos.
 
\item The probability distribution of $\cos\theta$ (i.e. the angle
  between the major axis $\Vec{E}_a$ of a subhalo and its radius
  vector in the reference frame of the host) can be well described by a
  simple power law $P(\cos\theta) \propto \cos^4\theta$

\item These results do not depend on the mass of the host system.

\end{itemize}

\section*{Acknowledgements}
AK and ND acknowledge funding through the Emmy Noether programme of
the DFG (KN 755/1). CP is supported by the STFC rolling grant for
theoretical astrophysics at the University of Leicester. The numerical
simulations of G1-5 and MW2-4 were performed at the LRZ Munich, NIC
J\"ulich and BSC Barcelona. We thank DEISA for providing time to these
computers under DECI project SIMU-LU. MC1-3, C1-12, and MW5 were
carried out on Swinburne University's supercomputer. The analysis was
partly performed on the Sanssouci cluster of the AIP and at the LRZ
Munich.


\bibliography{papers} \bsp

\label{lastpage}

\end{document}